# Artificial Neural Network based on SQUIDs: demonstration of network training and operation

**Chiarello F [1], Carelli P [2], Castellano M G [1], Torrioli G [1]**

[1] IFN-CNR, via Cineto Romano 42, 00156 Rome, Italy.
[2] DSFC, Università dell'Aquila, via Vetoio 1, 67100 L'Aquila, Italy

Fabio.chiarello@ifn.cnr.it

**Abstract**. We propose a scheme for the realization of artificial neural networks based on Superconducting Quantum Interference Devices (SQUIDs). In order to demonstrate the operation of this scheme we designed and successfully tested a small network that implements a XOR gate and is trained by means of examples. The proposed scheme can be particularly convenient as support for superconducting applications such as detectors for astrophysics, high energy experiments, medicine imaging and so on.

## 1. Artificial Neural Networks

Artificial Neural Networks (ANN) are a computing paradigm inspired by biological neural systems [1,2]. They are particularly effective in specific tasks such as pattern recognition, data mining, regression and analysis of series, classification, filtering and so on. An important characteristic of ANN is their capability to learn and to be trained in order to perform some required task. The training can be done once and for all during an initial preparatory phase (which can also be the design phase) if the network is used only for recurrent tasks, or it is possible to have a network that can learn during its use, in order to adapt itself to a changing environment.

ANN can be realized in many different ways according to the needs of specific applications: for example they can be simulated by computer programs, or implemented by discrete electronic circuits, by microcontrollers, DSPs, ASICs and so on [3,4]. Unconventional electronics can also be used for this task, for example there are different proposals based on the use of superconducting electronics [5–8]. Superconducting electronics allows computing speeds not possible with conventional semiconductor electronics (with clocks of the order of hundreds of Gigahertz) [9–11], but its drawback is the necessity to use cryogenic systems, with relative costs and request of competences. However there are different examples of important existing applications based on low temperatures and superconductivity (astrophysical detectors, medical imaging, high energy physics, quantum communication and computing, ultra low noise electronics, metrology and so on) where the introduction of a superconducting electronics is not cause of extra cost, but rather it could provide a natural interface between the superconducting system and the room temperature electronics, performing preliminary tasks such as multiplexing, pre-analysis and triggering [12–15]. In many cases the availability of a fast superconducting ANN close to the system could be a very useful and desirable tool.

In the present work we consider a simple and flexible scheme for the realization of ANNs with SQUIDs (Superconducting Quantum Interference Devices), a widely used class of superconducting devices characterized by outstanding performances as ultrasensitive magnetometers and low noise amplifiers [16–18]. In order to prove the operation of the proposed scheme we realized and tested a simple yet non trivial network, and demonstrated its training for a specific tasks. The paper is organized as follows: section 2 is a short introduction to ANNs for our purposes. In section 3 we describe the use of SQUIDs as artificial neurons. In section 4 experimental setup and results are discussed. Section 5 presents conclusions and remarks.

**2. Artificial Neural Networks**

An ANN can be represented by a graph where nodes are "neurons" (in fig.1a, circles numbered from 1 to 5) and connections are "synapses" (arrows in fig.1a).

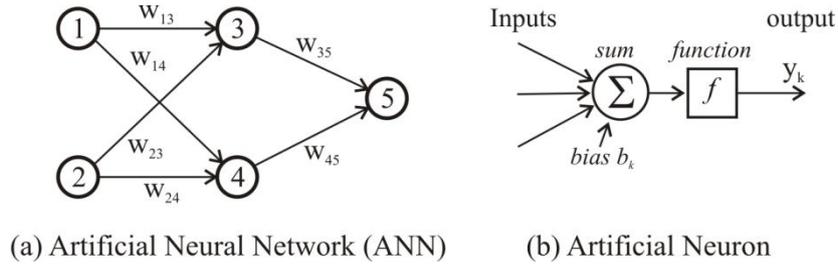

(a) Artificial Neural Network (ANN)    (b) Artificial Neuron

Fig.1. (a) Example of an artificial neural network. (b) Scheme of a single neuron.

A generic neuron $k$ is described by its internal state $x_k$, a real value that can be discrete or continuous depending on the particular problem considered. A synapse connects two neurons, for example $l$ and $k$, and has the effect to apply to neuron $k$ the state of neuron $l$ multiplied for a weight $w_{lk}$ (a real number). Each neuron acts as a primitive function, which sums together all the weighted inputs plus a bias $b_k$, and uses an activation function $f$ to evaluate the new state $y_k$ (fig.1b), the output value:

$$y_k = f\left(\sum_l w_{lk} x_l + b_k\right). \tag{1}$$

The activation function depends on the considered problem, and can be a step function, a sigmoid, a linear function and so on. Once an appropriate architecture has been chosen (number of neurons, activation functions, network topology and so on), the network must be trained in order to learn to perform the required task. This is done by properly modifying weights $w_{lk}$ and biases $b_k$ in order to have correct responses to applied input patterns. The training can be done once and for all in an initial design phase, if the network must do a recurrent task previously analyzed, or it can be done "on the run", for an adaptation to a variable environment. In principle an ANN must be able to perform all the tasks possible with a standard computer and this is possible if all the basic boolean operations can be done by neurons. A single artificial neuron can be immediately used as NOT, AND and OR gates, but the realization of a XOR gate is a bit more complicate and it has been a great problem for a long time in the history of artificial neural networks. The solution was found by introducing a multilayer topology of the kind presented in fig.1a, with three distinct layers: an input layer (neurons 1 and 2), a hidden layer (neurons 3 and 4), and an output layer (neuron 5). In order have a demonstrator at the same time simple (with no useless complications) but non-trivial (able to demonstrate the universality of the scheme), we chose to implement exactly the critical XOR gate. The simple ANN presented in fig.1a can be used as a XOR gate by considering neurons with discrete states (for example -1 and +1) and step activation functions, with weights $w_{13}= w_{24}= w_{35}= w_{45}= +1$, $w_{14}= w_{23}= -1$, and biases $b_1=b_2=0$, $b_3= b_4=B1<0$, $b_5=B2>0$. If we indicate with I1 and I2 the states of neurons 1 and 2, the new states of neurons 3 and 4 are given by $x_3 = sign(I1-I2+B1)$ and $x_4 = sign(-I1+I2+B1)$, where the function *sign(x)* returns -1 for $x<0$ and +1 for $x>0$. The total input at neuron 5 is the sum of these states, and it is simple to verify that this is equal to 2 for $|I1-I2|<B1$ and zero elsewhere when B1>0 (the result is inverted for B1<0). By setting B2=1 we can use neuron 5 to discriminate these states, obtaining a final output +1 if $|I1-I2|<B1$, and -1 elsewhere (or the opposite for B1<0). By assuming B1= -1 we obtain the desired XOR operation. This network is very robust against errors and fluctuations, and can continue to operate correctly also for large variations of weights, biases and input values.

Moreover, the network can perform a more complex task that the simple XOR gate by considering continuous states: it is able to recognize if the difference of the two inputs I1-I2 is in modulus smaller (or greater) than a given value B1. The network can learn this task directly from a series of examples, and can be trained, for example, by means of a genetic algorithm that modifies the biases $b_3$ and $b_4$ of the hidden layer neurons.

### 3. SQUID Artificial Neural Network

We present the use of a superconducting electronics based on SQUIDs and on magnetic flux signals for the implementation of general ANNs, with a specific example used as a demonstrator. In fig.2 it is shown the scheme of this example, corresponding to the network given in fig.1a.

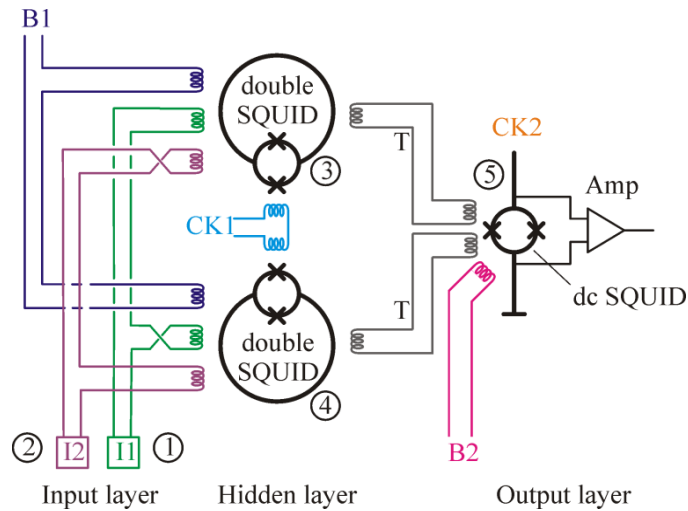

Fig.2. Scheme of the SQUID neural network, with current generators as input layer neurons 1 and 2, double SQUIDs as hidden layer neurons 3 and 4, and a dc SQUID as output layer neuron 5.

A SQUID consists of a superconducting loop interrupted by one or more Josephson junctions [16–18]. Typically a SQUID responds in a nonlinear way (by changing the total magnetic flux threading its ring) to the sum of the input magnetic fluxes applied to its loop. We use SQUIDs as neurons, and magnetic fluxes as states. Different neurons can be linked together by means of superconducting transformers, which are closed superconducting loop inductively coupled to the SQUID rings (fig.2, see the transformers T between neurons 3 and 4 in the hidden layer, and neuron 5 in the output layer). The total flux in a superconducting loop is quantized and cannot change (it must be an integer number of flux quanta, $\Phi_0 = h/(2e)$, and this number cannot be modified unless a transition to the normal state). Therefore, if there are one or more external fluxes $\Phi_k$ applied to the ring, a supercurrent $I$ must flow in the loop in order to compensate the sum $\sum \Phi_k$ and so maintain the total flux constant. The effect of a flux applied in any point is transported along all the loop, and this can be used in order to magnetically connect devices in different positions, with different fixed weights (determined by the coupling strength) and fixed versus(determined by the direction of the couplings).

An underdamped dc SQUID (direct-current SQUID) consists of a superconducting loop of inductance $l$ interrupted by two (nominally) identical junctions of capacitance $c$ and critical current $i_0$, biased by a current *I* (fig.2, in the output layer). For small inductance ($l \ll \Phi_0 / i_0$) the device behaves approximately like a single Josephson junction with total capacitance *C=2c* and critical current $I_c(\Phi_c) = 2i_0 \cos(\pi \Phi / \Phi_0)$, which is controlled by the sum of all the magnetic fluxes $\Phi_c$ applied to the ring. For *I<I$_c$* the device remains in the superconducting state and the voltage at its electrodes is zero, but when *I>I$_c$* there is a transition to the voltage state and a voltage $V \approx 2.7\,mV$ (for Niobium junctions) appears. Since $I_c$ depends on the total flux applied, this device is often used as magnetometer for measuring or discriminating magnetic fluxes. In our case the dc SQUIDs is used as a neuron of the output layer. The fluxes coming from input neurons are summed together with an extra bias flux (B2) in the SQUID loop. When required, a current ramp is applied to the device (CK2) and the critical current at which a voltage transition occurs is recorded. The total input flux is determined by this value and it is possible to apply the activation function. In this kind of neuron the input quantities are magnetic fluxes while at the output there are voltages/currents (which are simple to be recorded). This is ideal for the final stage (output layer), where it is necessary to read out the neuron state, but cannot be used for intermediate neurons, where both inputs and outputs must be magnetic fluxes.

For the intermediate layer we can use a double SQUID, a superconducting loop of inductance *L* interrupted by a dc SQUID (fig.2, neurons 3 and 4). In this case the output state, the total flux $\Phi$ threading the large loop, changes in a non linear manner with the sum $\Phi_x$ of the magnetic fluxes applied to it (input states), and this behavior is controlled by the flux applied to the small loop (CK1). The device can be described by an equivalent mechanical model, with an effective mass $M = C\Phi_b^2$ (where $\Phi_b = \Phi_0 / (2\pi)$ is the reduced flux quantum) moving along a fictitious direction $\Phi$ in an effective potential [19]:

$$U = \frac{(\Phi - \Phi_x)^2}{2L} - I_0 \Phi_b \cos\left(\frac{\Phi}{\Phi_b}\right), \qquad (2)$$

where the critical current $I_0$ is modulated by the flux applied to the small loop (CK1). If the changes are slow enough (adiabatic) the system always remains in a minimum, so that the device response is the flux $\Phi_m$ corresponding to a minimum, modulated by the applied input flux $\Phi_x$. The position of minima is given by solving the implicit equation $\Phi_x = \Phi_m + I_0 L \sin(\Phi_m / \Phi_b)$, which is periodic for translations $\Phi_x \to \Phi_x + n\Phi_b$ and $\Phi_m \to \Phi_m + n\Phi_b$ (where *n* is an integer). When $I_0 < \Phi_b / L$ the relation is single-valued and for $I_0 = 0$ it is linear (corresponding to a neuron with a linear activation function), while for $I_0 \to \Phi_b / L$ it is strongly nonlinear, like a step function for $\Phi_x \approx \Phi_0 / 2$. For $I_0 > \Phi_b / L$ the function is multi-valued and the presence of more minima can be exploited in order to introduce a memory and use the device as a flip-flop. A temporary reduction of $I_0$ below $\Phi_b / L$ (controlled by the flux CK1) has the effect of a reset. The return of $I_0$ above $\Phi_b / L$ causes the retrapping of the system in one of the minima, memorizing the state of the input flux $\Phi_x$. This effect can be used in order to realize neurons with memory and with the possibility to be synchronized by a clock signal (the control flux CK1). The synchronization is important in order to ensure the forward propagation of the information from the input to the output.

Both dc SQUID and double SQUID are affected by noise, in particular by thermal noise, and their behavior is subjected to errors. For example, in the use of a SQUID as a step function the transition between opposite states is not as sharp as in the ideal case and there is a gray zone where the result is not deterministic. In our test system we have replaced the loops of all SQUIDs with the parallel of couple of identical loops. This gradiometric configuration allows to reduce the pick-up of flux noise and simplify the design of coupling between neurons.

A device similar to the double SQUID, the tunable transformer, can be used in order to have tunable weights, allowing the ANN to learn "on the run". In this case a dc SQUID is inserted in a superconducting transformer, realizing a double SQUID with two large loops in parallel. The dc SQUID is used as a switch that can short-circuit the transmission of the signal from one loop to the other. This device has been extensively studied in the past [20], and could be integrated in the present test circuit.

In our demonstrator neurons 1 and 2 in the input layer are implemented by currents flowing in coils. In a network developed for a real applications these neurons must ensure the interface towards the system external to the network. For example, if the network is used to analyze or multiplex current signals coming from astrophysical detectors, the ideal input neurons are double SQUIDs, which allow amplification, readout and memorization of the input signals.

It is interesting to note that our SQUID neurons have periodic behaviors with respect to their flux inputs, so that also their activation function is periodic. It is possible to have linear or step activation functions by limiting the input range within a period but, if required, it is also possible to exploit the periodicity for more flexible tasks.

These SQUID neurons have a large gain in the input – output characteristics. A small input magnetic flux, just large enough to overcome the gray zone due to noise (typically of the order of 10 $m\Phi_0$ at 4.2 K), may cause the jump of one flux quantum in the SQUID output state. This corresponds to a gain of about 100.

## 4. Experimental setup, results and discussions

We have designed a series of chips based on the scheme in Fig.2, which have been fabricated at the MIT Lincoln Laboratory facility by using the deep submicron (DSM) with fully-planarized Nb-(Al-AlOx)-Nb trilayer process (fig.3).

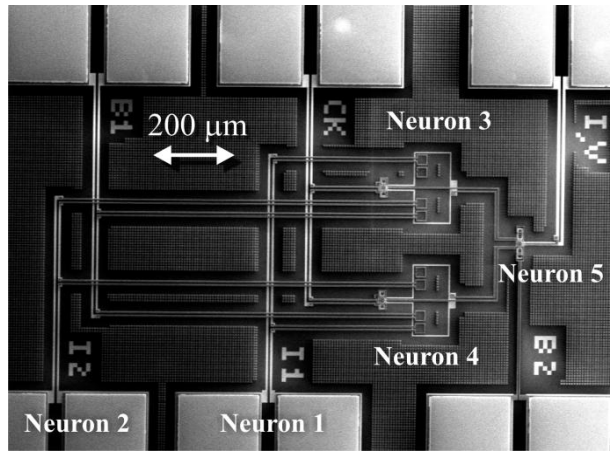

Fig.3. SEM photograph of the ANN under test. There are visible the three SQUIDs (neurons 3, 4 and 5), with gradiometric configurations (doubled loops), the superconducting transformers T, and the different bias lines.

The system has been tested at IFN-CNR Rome in a 3He cryostat "Heliox" with temperatures from 4.2K down to 350mK (the lower temperature is used in order to reduce thermal noise and so to improve the sharpness of the step function). The nominal parameters of the dc SQUID are $i_0 \cong 8\ \mu A$, $c \cong 10\ fF$ and $l \cong 10\ pH$; for the double SQUIDs we have identical values (for the inner dc SQUID), and a large inductance $L \cong 230\ pH$. The network is fed by two bias currents (B1 and B2), by two input currents (I1 and I2), and by a couple of clock signals (CK1 and CK2).

First of all, the dc SQUID (neuron 5) can be characterized by observing its interference pattern (fig.4), obtained by applying current ramps to CK2 and observing the variation of the critical current for different constant B2. The anomalous bumps on minima in fig.4 are a spurious instrumental effect.

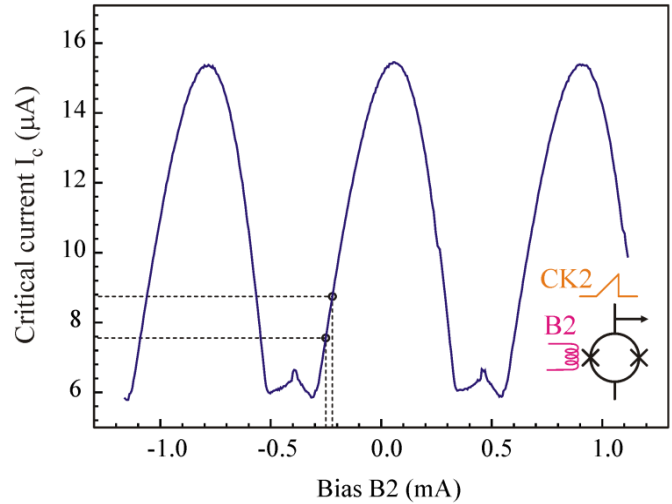

Fig.4 Experimental characterization of the dc SQUID (schematized in the inset on the left), obtained by evaluating the modulation of the critical current $I_c$ for different bias values B2. Dashed lines indicate a typical operating region.

Then the dc SQUID is used for the readout of the full system, considering crossed configurations of the different bias currents applied to I1, I2, B1, CK1, and taking into account the periodicity in the behaviour of the SQUIDs. We measure $M_{B2}$ = 2.5 pH for the coupling between "B2" and the dc SQUID, $M_{I1}$ = 4.7 pH, $M_{I2}$=6.4 pH and $M_{B1}$=6.2 pH for the couplings between coils I1, I2 and B1 and the relative large loops in the double SQUIDs, $M_{ck1}$ = 3.8 pH for the coupling between CK1 and the relative small loops in the double SQUIDs. We measured a transforming ratio of 0.7% for the coupling between the double SQUIDs and the dc SQUID. All these values are in good agreement with the design parameters (except for $M_{I1}$, which is lower). These characterizations will allow later optimizations for the values applied to B1 and B2 and for the baselines and amplitudes of the clocks CK1 and CK2.

In order to demonstrate the correct operation of the network as a XOR gate we optimize the amplitudes of biases B1 and B2, and introduces a further bias B0 summed to the input I1 that allows to compensate possible trapped fluxes. A clock sequence is applied to CK1 and CK2 with a repetition rate of 2kHz, formed by a first short pulse (80μs) on CK1, used for the reset/set of the neurons in the hidden layer, followed by a current ramp on CK2, starting after 200μs from the first pulse. The ramp moves the current above the critical value, which is detected and recorded thanks to the transition to the voltage state. This procedure is repeated for different input values I1 and I2 in order to obtain the input – output characteristic of the network. Fig.5 shows the measured output critical current $I_c$ vs. the two input currents applied to I1 and I2, plotted as a density map.

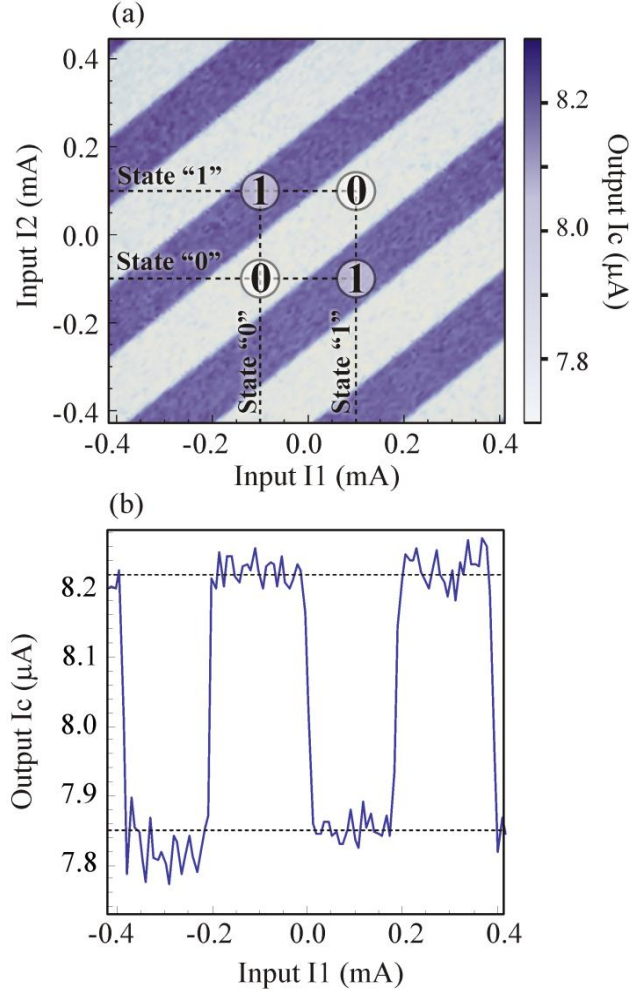

Fig.5 (a) Experimental density map showing the measured response of the network to different input currents B1 and B2. Circles and dashed lines illustrate the use of the network as a XOR gate. (b) Output critical current obtained for fixed I2 = 0.1 mA.

The XOR operation is verified by associating the logical states "0"/"1" to currents +0.1mA/-0.1mA in both I1 and I2 inputs, and 7.8μA/8.2μA for the output value. We observe large regions where the system works correctly (see circles in fig.5), indicating the robustness of the network for variations of the input values. Note also the periodicity in the input currents. In this case we demonstrated the operation of the network with no training or, to be more precise, with a training done at the design phase.

Let us present a second case, with a running training based on examples. By considering continuous quantities for the states I1 and I2, the network is able to verify the relation $\mod\left[M_{I1}(I1+B0) - M_{I2}I2\right] < M_{B1}B1$ (where the function mod(x) stands for modulus $\Phi_0$). This means that the network is able to discriminate if a point in the I1-I2 plane is inside a dark or a light oblique band in fig.5 (in that figure B1 determines the relative width of bands and B2 the shift in the I1 direction). Now we firstly set our problem by fixing a couple of ideal values $B0^*$ and $B1^*$. We generate a set of example patterns (400 examples) by choosing randomly couples of inputs I1 and I2 and calculating the relative expected output from the ideal relation and the chosen biases $B0^*$ and $B1^*$. Then we start the training of the network, which must learn to do the correct discrimination from the set of examples. For this training we use a simple genetic algorithm: we consider 20 "individuals", each of them characterized by a "genome" (a couple of values B0, B1), initially assigned randomly. A set of 10 examples, randomly chosen between the set of example patterns, are presented to the (real) network, and this is repeated for each individual (for each couple of values B0 and B1) recording the number of obtained successes. The five individuals with higher number of successes are maintained, while the other 15 are replaced by replicas of the best five with small random modifications in their genome. This operation produces a new generation of individuals, more skilled in the required task. This is repeated for 100 generations. At the end we obtain 20 individuals with parameters B0 and B1 very close to the ideal values B0$^*$ and B1$^*$, and chose the best of them to fix the trained network. Fig.6 displays the percentage of failures obtained after each generation, with a value starting from 50% for the first generation (which is randomly chosen), and saturating at 4%. The nonzero value of this final failure-rate is due to the presence of noise, which causes a non sharp transition between the bands described in fig.4, with a "gray zone" where the discrimination is uncertain. We repeated the training for different initial ideal values B0$^*$ and B1$^*$ and different sets of examples obtaining very similar results, but observing that the 4% value is the best (lowest) possible, obtained in the case of not too narrow bands. If one of the two kinds of bands is too narrow (comparable with the gray-zone width) the discrimination becomes more and more difficult and the failure-rate grows.

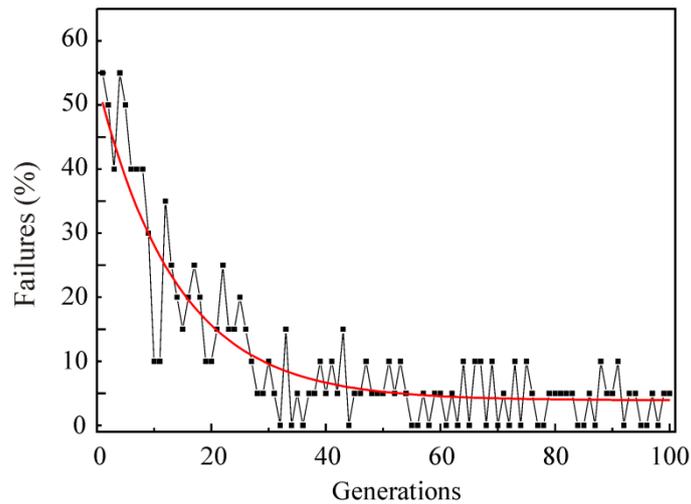

Fig.6. Network training: experimental numbers of failure for each new generation of the genetic algorithm.

After the training phase the network must be validated. A new set of 400 ideal examples are randomly generated and applied to the network with optimal parameters. Also in this case we obtain a best failure rate of about 4%, as expected. We repeated the training and verification of the network at lower temperature (350mK) in order to reduce the thermal noise, and in the same conditions we obtain a reduced failure rate of 2%, as expected. From an "a posteriori" measurement and evaluation of the noise effects on the system, we conclude that it would be possible to drastically reduce this failure-rate by fabricating a new network with increased couplings between double SQUIDs (neurons 3 and 4) and the dc SQUID (neuron 5).

All the measurements are performed at a repetition rate of 2kHz. This speed is only due to our slow electronics and it is not an intrinsic limit of the network. In order to estimate the real speed we observe that in our scheme the operations of double SQUIDs are based on adiabatic modifications of their energy potentials (where a timescale $\sqrt{LC} = 2.14\,ps$ dominates), while the operations on the dc SQUIDs are based on transitions between superconducting and voltage states (with a typical timescale $RC \sim 100ps$). A simple simulation shows that the maximum rate for the correct operation must be of the order of few GHz. This time can be considerably increased by introducing dissipative elements, in particular by shunting the Josephson junctions with small resistances, in analogy with what is done in conventional superconducting electronics (for example, for Rapid Single Flux Quantum electronics). In this case we expect possible rates up to hundreds of GHz [10,11].

It is interesting to notice that all the elements used in our scheme (the different SQUIDs and the superconducting transformers) are the same diffusely used in quantum computing applications as qubits (and relative readout and coupling "accessories"), with very similar configurations and parameters. Here we proposed and tested systems operating only in the classical regime, but there are no real limits to their use as quantum devices. Some authors suggest the possibility to introduce quantum logic in ANNs in order to extend their capabilities, similarly to what happens with the introduction of quantum computing [21,22]. In this direction, our scheme could be suitable for the real implementation of a possible quantum ANN.

## 5. Conclusions

We proposed a scheme for the implementation of artificial neural networks based on SQUIDs, and demonstrated its operation thanks to a simple test system consisting of two double SQUID and a dc SQUID, coupled by means of superconducting transformers. The system has operated correctly as a XOR gate and has been also used for a more complex task, demonstrating the feasibility of training based on examples. The presented scheme is particularly convenient in concurrence with superconducting devices used for other applications, such as detectors for astrophysics, high energy experiments, medical imaging and so on, where SQUIDs are well suited for the device readout, and the use of cryogenic technologies is not a limit. In case of a possible future development in the field of quantum ANNs, the presented scheme could be a suitable candidate for a practical implementation of such systems.

**Acknowledgments**
This work was supported by Italian MIUR under the PRIN2008 C3JE43 project. We tank W. Oliver and the MIT Lincoln Laboratory LTSE team for device fabrication, S. Cibella for useful help and M. Caretti for discussions.